\documentclass[]{spie}  
\pdfoutput=1 
\usepackage[]{times}
\usepackage[]{graphicx}
\usepackage{amsmath}
\usepackage{amssymb}
\usepackage{hyperref}
\usepackage[seperr]{siunitx}
\provideunit{\ph}{photon}
\provideunit{\pix}{pixel}
\provideunit{\tarcsecond}{arcsecond}
\provideunit{\el}{e\tothe{-}}
\provideunit{\psb}{\ph\per\second\per\Square\metre\per\Square\tarcsecond\per\micro\metre}
\provideunit{\micron}{\micro\metre}
\title{Photonic lantern behaviour and implications for instrument design} 

\author{Anthony Horton\supit{a}, Robert Content\supit{a}, Simon Ellis\supit{a} and Jon Lawrence\supit{a}
\skiplinehalf
\supit{a}Australian Astronomical Observatory, PO Box 915, North Ryde NSW 1670, Australia
}

\authorinfo{Further author information (send correspondence to AJH):\\AJH: E-mail: \href{mailto:anthony.horton@aao.gov.au}{\nolinkurl{anthony.horton@aao.gov.au}}, Telephone: +61 2 9372 4847\\}

\begin{document} 
  
Anthony Horton, Robert Content, Simon Ellis and Jon Lawrence, "Photonic lantern behaviour and implications for instrument design," {Advances in Optical and Mechanical Technologies for Telescopes and Instrumentation, Navarro, Cunningham, Barto, Editors}, {\em Proc.~SPIE}~{\bf 9151}, 9151-72 (2014)

Copyright 2014 Society of Photo Optical Instrumentation Engineers. One print or electronic copy may be made for personal use only. Systematic electronic or print reproduction and distribution, duplication of any material in this paper for a fee or for commercial purposes, or modification of the content of the paper are prohibited.

\url{http://dx.doi.org/}

\pagebreak
\mbox{}
\pagebreak

  \maketitle 

\begin{abstract}
Photonic lanterns are an important enabling technology for
astrophotonics with a wide range of potential applications including
fibre Bragg grating OH suppression, integrated photonic spectrographs
and fibre scramblers for high resolution spectroscopy.  The behaviour
of photonic lanterns differs in several important respects from the
conventional fibre systems more frequently used in astronomical
instruments and a detailed understanding of this behaviour is required
in order to make the most effective use of this promising technology.
To this end we have undertaken a laboratory study of photonic lanterns
with the aim of developing an empirical model for the mapping from
input to output illumination distributions.  We have measured overall
transmission and near field output light distributions as a
function of input angle of incidence for photonic lanterns with
between 19 and 61 cores.  We present the results of this work,
highlight the key differences between photonic lanterns and
conventional fibres, and illustrate the implications for instrument
design via a case study, the design of the PRAXIS spectrograph.  The
empirical photonic lantern model was incorporated into an end-to-end
PRAXIS performance model which was used to optimise the design parameters of
the instrument.  We describe the methods used and the resulting
conclusions.  The details of photonic lantern behaviour proved
particularly important in selecting the optimum on sky field of view
per fibre and in modelling of the instrument thermal background.
\end{abstract}


\keywords{Astrophotonics, photonics, photonic lantern, optical fibre, fibre Bragg grating, OH suppression, spectroscopy}

\section{INTRODUCTION}
\label{sec:intro}  

In recent years there has been considerable interest in the field of
astrophotonics, the application of photonic technologies to
astronomy\cite{Bland-Hawthorn2009a}.  Photonic devices such as
fibre Bragg gratings (FBGs), arrayed waveguide gratings (AWGs) and
integrated optics interferometers offer considerable benefits in a
range of astronomical applications, specifically OH suppression
(OHS)\cite{Ellis2008,Bland-Hawthorn2011},
spectroscopy\cite{Cvetojevic2009,Allington-Smith2010} and
optical interferometry\cite{LeBouquin2011} for the devices listed.  One
property shared by all of these devices is that their operation
depends critically on their being constructed from single mode fibres/waveguides
and this presents a challenge for astronomical use.  The
coupling efficiency between a telescope and a single mode
fibre is approximatedly 0.7 times the Strehl ratio of the stellar
image, for reasonable values of the telescope central
obstruction\cite{Shaklan1988,CoudduForesto2000}. Under some
circumstances the resulting coupling losses are acceptable, i.e.\ 
narrow field adaptive optics (AO) on telescopes that are either small or
have high order (`extreme') AO systems, and in the case of interferometry
the rejection of light by the single mode fibre actually acts as
beneficial spatial filter.  In general, though, direct coupling into
single mode fibres is unacceptably lossy, for example for seeing limited
observations on large telescopes the Strehl ratio and consequently the
coupling efficiency will be well below 1\%. The photonic lantern was
invented as a solution to this problem.

A photonic lantern is a fibre (or waveguide) taper device which converts
a single multimode input into multiple single mode outputs or vice
versa.  With appropriate design parameters and the right
illumination conditions the transition can be low loss in both
directions\cite{Noordegraaf2009,Leon-Saval2010}.  By using a photonic
lantern to feed multimode light into an array of identical single mode
photonic devices and (optionally) a second photonic lantern to combine
the light from their outputs into a single multimode output it is
possible to construct a multimode device with the same performance of
a single mode device, albeit at the cost of producing multiple copies
of the single mode device\cite{Leon-Saval2005a}.  As a result photonic
lanterns are a key part of the practical implementation of most
astrophotonic technologies.

Photonic lanterns also have potential astronomical applications of
their own.  Experimental results suggest that photonic lanterns are
highly effective `fibre scramblers', i.e.\ the output illumination
distribution is insensitive to the input illumination
distribution\cite{Olaya2012a}.  This would make photonic lanterns
useful for stabilising the point spread functions of spectrographs,
such as those used for high precision radial velocity measurements.

In order to utilise photonic lanterns effectively it is important to
understand their behaviour.  This behaviour differs both qualitatively
and quantitively from the conventional step index multimode fibres
(MMF) typically used for astronomy, primarily because waveguide modal
behaviour manifests differently in photonic lanterns.  
In a conventional MMF the number of
guided modes that can propagate along the fibre is
determined by the material numerical aperture
($\textrm{NA}=\sqrt{n_\textrm{core}^2 - n_\textrm{clad}^2}$ where
$n_\textrm{core}$ and $n_\textrm{clad}$ are the refractive indices
of the fibre core and cladding materials respectively), the fibre core
diameter ($d$) and the wavelength ($\lambda$) via the equation
\begin{equation}
\label{eqn:n_modes}
N_\textrm{modes} \approx \frac{\left(\pi d \textrm{NA}\right)^2}{2\lambda^2}.
\end{equation}
For a photonic lantern, on the other hand, the number of modes which
can propagate from the multimode input to the multimode output is
fixed and equal to twice the number of single mode
fibres/cores/waveguides within the lantern (hereafter referred to as
cores).  Equation \ref{eqn:n_modes} is still valid
however, and can be rearranged to define the effective numerical aperture
of a photonic lantern\cite{Noordegraaf2012},
\begin{equation}
  \label{eqn:napl}
  \textrm{NA}_\textrm{PL} \approx \frac{2\lambda \sqrt{N_\textrm{core}}}{\pi d}.
\end{equation}
NA is an important design parameter as it defines the acceptance cone
angle for the multimode input of the photonic lantern, $\textrm{NA} =
n\sin\theta$ where $\theta$ is the half cone angle and $n$ is the
refractive index of the medium in front of the input. By applying
conservation of \'{e}tendue/$A\Omega$ product and using small angle
approximations we are able to write a simple expression relating the
on sky field of view (FoV) per photonic lantern of an astronomical
instrument to the number of cores in each lantern,
\begin{equation}
\label{eqn:fov}
\theta_{\textrm{sky}} \approx \frac{4\lambda \sqrt{N}}{\pi D \phi_{\textrm{in}}},
\end{equation}
where $\theta_{\textrm{sky}}$ is the angular diameter of the lantern FoV, $D$ is the
diameter of the telescope and $\phi_{\textrm{in}}$ is a parameter
$\ge 1$ to
account for focal ratio degradation (FRD) in any sections of MMF
on the input side of the photonic lantern.

Naively we might assume that equations \ref{eqn:napl} and
\ref{eqn:fov} define the optimal NA with which to illuminate a
photonic lantern and the corresponding optimal on-sky FoV per lantern,
however to do so is to make an implicit assumption that all light with
an NA below the photonic lantern NA will be transmitted and all light
with an NA above the photonic lantern NA will be lost.  In reality the
situation is not so clear cut.

For a conventional step index MMF with a very large number of guided
modes the qualitative behaviour can be deduced from consideration of geometric
(ray) optics.  Any light entering the fibre with an NA less than the
material NA of the fibre will be totally
internally reflected at the core-clad interface and guided by the
fibre.  Light which is incident at larger angles may or may not be
reflected at the core-clad interface depending on its position. If a
ray is laterally offset from the fibre axis (a `skew ray') then it
will intersect the core-clad interface at an oblique angle and if the
offset is large enough it will be internally reflected.  As a result
the transmission of the fibre as a function of the incident angle of
illumination is high and constant until the angle reaches the fibre's
material NA at which point it tails off until only the most extreme
skew rays are guided by the fibre.  A full waveguide analysis only
slightly alters these conclusions.  Skew rays beyond the material NA
of the fibre are found to correspond to `leaky modes' which radiate
energy as they propagate along the fibre in a process analogous to
bend losses.  As a consequence the transmission of the fibre beyond
the material NA reduces as fibre length increases so that for an
infinitely long fibre the behaviour reduces to a sharp cut off at the
material NA. 

In the opposite limit, that of a single mode fibre (SMF), the coupling of
light into the fibre is determined by the overlap integral of the
incident electromagnetic field with the fibre mode field.  The mode
field of a step index single mode fibre is approximately Gaussian in
form and so fibre transmission as a function of incident angle of
illumination is also approximately Gaussian.

Photonic lanterns, which are able to propagate a small and fixed
number of modes, are expected to exhibit behaviour intermediate
between that of a highly multimoded fibre and a single mode fibre.
The exact nature of this behaviour is not well known, however, but is
important for determining the optimum design parameters for instruments
that use photonic lanterns.  It is this issue that has been the focus
of our laboratory study to date.

\section{Filled cone tests}
\label{sec:filledcone}

\subsection{Method}

GNOSIS was the first instrument to use photonic lanterns for
astronomical observations\cite{Ellis2012,Trinh2013a}.  During the
commissioning of the GNOSIS instrument measurements were made of the
throughput of the OH suppression grating unit as a function of
input focal ratio.  These tests were performed on all 7 channels of the
complete grating unit.  Each channel consists of the following
components, in order:

\begin{enumerate}
\item{\SI{50}{\micro\metre} core diameter $\textrm{NA}=0.22$ MMF pigtail}
\item{A 1 MMF to 19 SMF (19 core) photonic lantern}
\item{19 H1 OH suppression FBGs}
\item{19 H2 OH suppression FBGs}
\item{A 19 SMF to 1 MMF (19 core) photonic lantern}
\item{\SI{50}{\micro\metre} core diameter $\textrm{NA}=0.22$ MMF pigtail}
\end{enumerate}

To perform the experiment the output of a fibre coupled
superluminescent diode (SLD) was collimated, passed through an
adjustable iris and then refocussed onto the input fibre pigtail.  An
infrared optical power meter was used to measure the power at both the
input and the output allowing the throughput to be calculated.

\subsection{Results}


\begin{figure}
\center\includegraphics[width=0.6\textwidth]{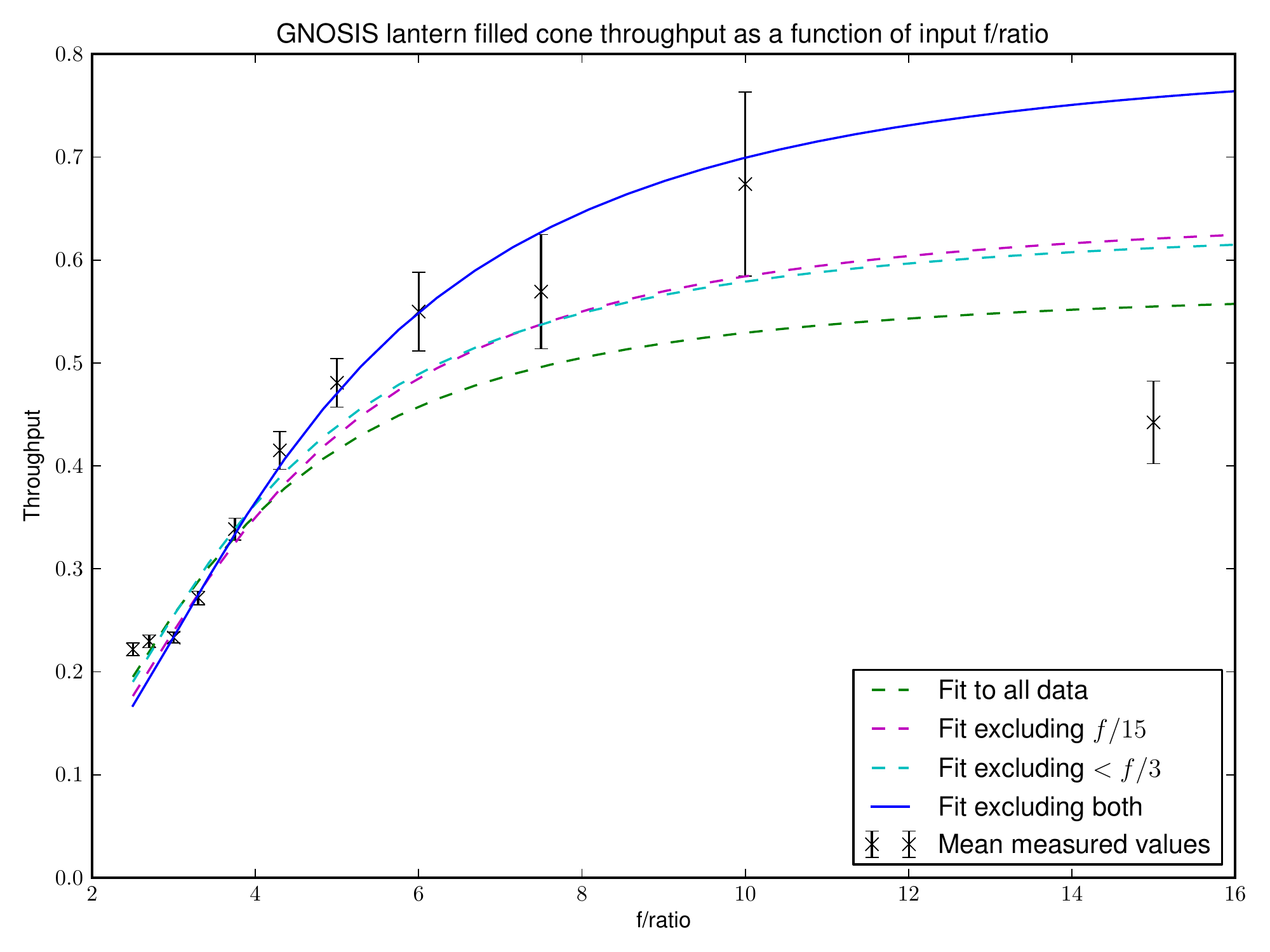}
\caption{\label{fig:gdata}
Measured throughput of the GNOSIS photonic lanterns when illuminated
by converging beams of various focal ratios.  The data shown are the
mean values at each focal ratio for the 7 channels of the GNOSIS
grating unit and are plotted together with the standard error on those
means.  Also shown are fits to the data using a Gaussian double
integral fitting function.}
\end{figure}

The results of the measurements of the GNOSIS photonic lanterns are
presented in figure \ref{fig:gdata} together with a series of fits to
the data.  We note that the measurements are generally consistent with a smooth
curve of increasing throughput with increasing focal ratio with the
exception of three data points, the two at lowest focal ratio and the
one at the highest focal ratio.  There are good reasons to believe
that systematic errors have effected these three values.  At the low
focal ratio end the higher than expected throughput values can be
explained by the collimated beam not fully or evenly illuminating the iris
aperture when it is at its widest, in effect the real focal ratio of
the illumination is lower than the diameter of the aperture would
suggest.  For the highest focal ratio point the lower than expected
throughput values can be explained by losses due to diffraction, at
$f/15$ the diffraction limited Airy disc diameter is
\SI{56}{\micro\meter}, larger than the \SI{50}{\micro\meter} diameter
core of the input fibre pigtail.

\subsection{Analysis}



As these photonic lanterns can only transmit a small number
of modes we expected their behaviour to be qualitatively similar to that
of a single mode fibre.  We therefore tried a fitting function which
assumes that the transmission of the lantern as a function of input
angle (or, equivalently, NA) is a Gaussian.  That function applies to
collimated illumination, to convert to illumination by a converging
beam as in this experiment we simply integrate the Gaussian over the
range of solid angles in the beam.  We performed a least squares fit
of the Gaussian double integral function to the complete data set, the
data set excluding the highest focal ratio point, the data set
excluding the two lowest focal ratio points, and the data set
excluding all three outliers.  As can be seen in the figure
\ref{fig:gdata} the fit is poor unless all three outliers are
excluded, in which case the simple Gaussian model reproduces the
measured results well.  

\section{Collimated illumination tests}

Due to time constraints during GNOSIS commissioning the filled cone
tests described in section \ref{sec:filledcone} inevitably provided
data limited in both quantity and quality.  The small number of data
points, large error bars and concerns about systematic errors mean
that the shape of the photonic lantern transmission function is poorly
constrained by these data.  In order to improve the state of
knowledge of the transmission function further tests of the GNOSIS
lanterns were necessary.  Furthermore the existing data on the GNOSIS
lanterns only provided information on the behaviour of a single
photonic lantern design with 19 cores, in order to gain insight into
the dependence of the shape of the transmission function on the number
of cores we have also undertaken tests on a 61 core lantern.

\subsection{Method}

\begin{figure}
\center\includegraphics[width=0.6\textwidth]{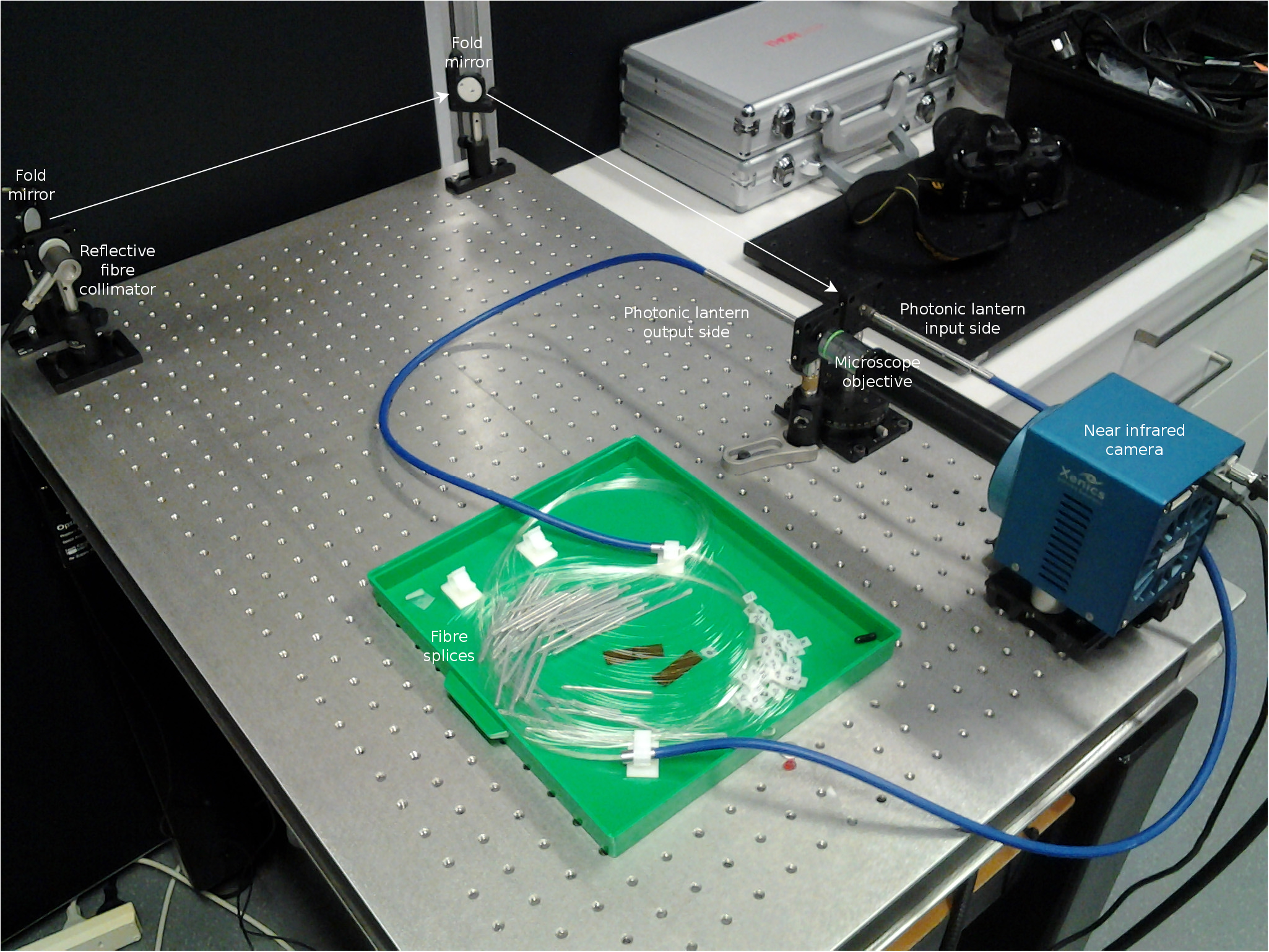}
\caption{\label{fig:lab_setup}
The collimated illumination photonic lantern test apparatus.  The
fibre coupled SLD source is connected to a reflective fibre collimator
and illuminates the input side of the photonic lantern via two
adjustable fold mirrors which were used for alignment.  The rest of the
apparatus can be seen more clearly in figure \ref{fig:lab_setup_closeup}.}
\end{figure}

\begin{figure}
\center\includegraphics[width=0.6\textwidth]{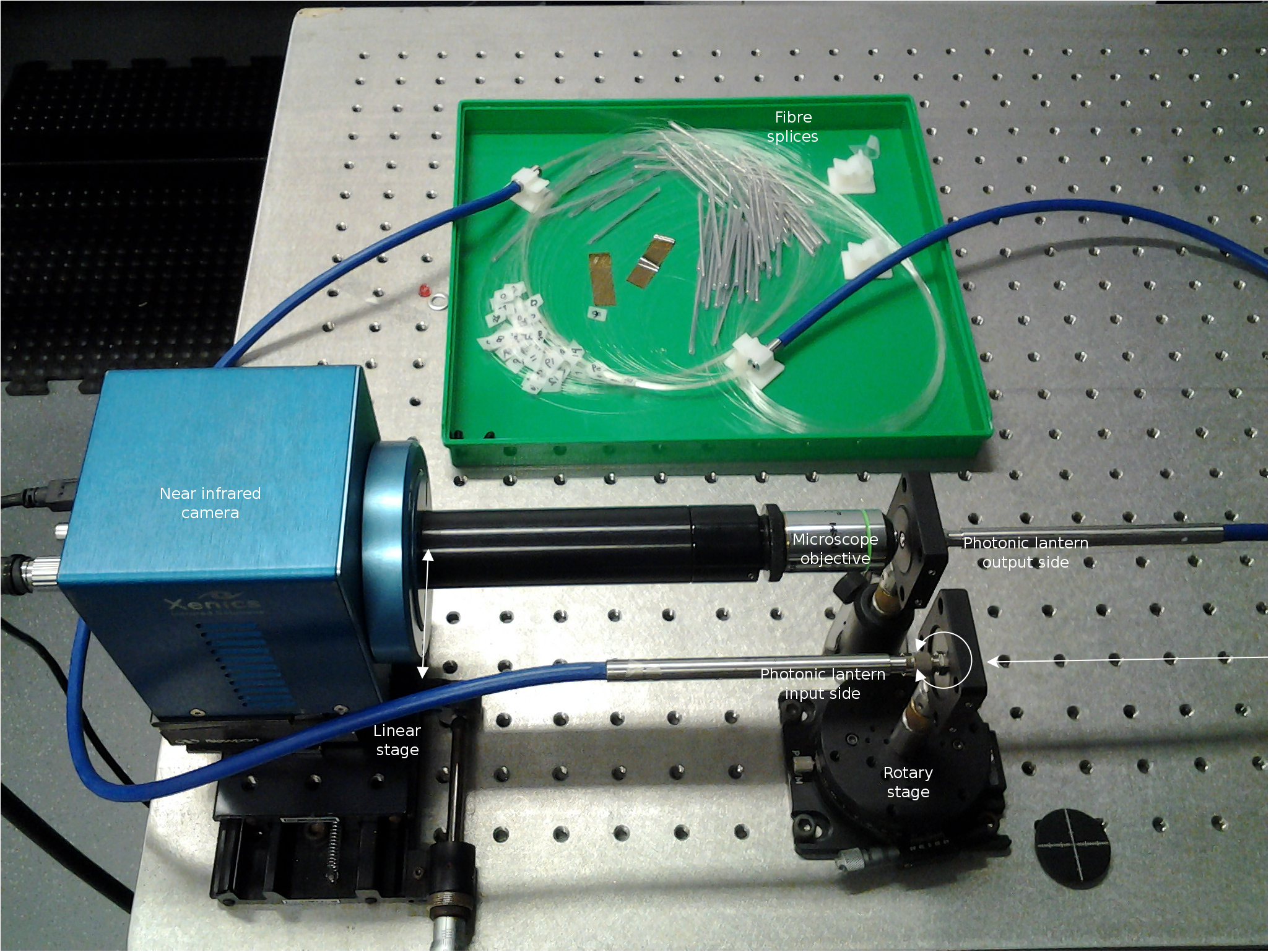}
\caption{\label{fig:lab_setup_closeup}
Close up of the collimated illumination photonic lantern test
apparatus.  The input side of the photonic lantern is mounted to a
rotary stage in such a way that the axis of rotation passes through
the end face of the lantern.  The output side of the photonic lantern
is mounted to a fixed post and the near infrared camera is used to
image the end face of the lantern (near field) through an
NA 0.4 microscope objective.  The near infrared camera is
mounted on a linear stage so that it can be moved across to image the input beam.}
\end{figure}

The experimental approach was somewhat different to that used during GNOSIS commissioning.  The same fibre
coupled SLD light source was used but instead of focussing the beam onto
the input of the photonic lantern under test we used collimated
illumination.  The input of the lantern was mounted in a rotary stage so that
it could be rotated about an axis perpendicular to the lantern axis
and passing through the input end face, thereby allowing the angle of
incidence of the illumination to be varied while keeping the photonic
lantern input in the same part of the beam.  Both converging and
collimated illumination are able to provide the same information, the
main advantages of the collimated approach are that it is less susceptible
to systematic errors caused by either uneven illumination of the beam
or diffraction.  One disadvantage of the collimated illumination
approach is that much less of the light from the source enters the
photonic lantern so more sensitive optical power measurements are
required, for this reason we used a near infrared camera to
measure the power instead of a photodiode based
optical power meter.  The experimental apparatus is shown in figures
\ref{fig:lab_setup} and \ref{fig:lab_setup_closeup}.

The GNOSIS lantern tests were performed on the single grating unit
channel which had FC/PC fibre connectors at input and output (the other 6
channels had their inputs and outputs permanently fusion spliced to
the fore optics unit and slit optics unit fibre bundles during
commissioning).  In order to connect the grating unit to the
experimental apparatus two Thorlabs M16L01 \SI{50}{\micro\metre} core, 0.22
NA, FC/PC to SMA \SI{1}{\metre} long fibre patch cables were connected
between the input and output ports of the experiment and those of the
grating unit.  The grating unit channel used for these tests is
believed to have the lowest throughput of the 7 based on test results
from GNOSIS commissioning.

The 61 core photonic lanterns used for this test were prototype
devices manufactured in 2009 by Crystal Fibre (now part of NKT
Photonics)\cite{Noordegraaf2010a}.  The photonic lanterns were supplied as packaged fibre
tapers with an SMA connector on the multimode side and bare single
mode fibre pigtails on the other.  In order to construct a complete
multimode to multiple single mode to multimode photonic lantern the 61
single mode fibre pigtails from one device were fusion spliced
directly to the 61 single mode fibre pigtails of a second device.  No
attempt was made to match up pigtails corresponding to the same
position within the fibre taper.  The two devices used have
designations PL12 and PL13. Both devices have nominal multimode core and cladding
diameters of \SI{102(5)}{\micro\metre} and \SI{130(10)}{\micro\metre}
respectively, and use Corning SMF-28 for the single mode pigtails.

For comparison with the photonic lanterns we also
tested conventional multimode fibres of similar core size.  
For the GNOSIS lantern the comparison fibres used were
simply the same two M16L01 fibre patch cords that had been used to connect
the lantern to the experiment, for the purposes of the comparison the
FC/PC connectorised ends of the two fibres were connected
directly to each other using a mating sleeve thereby bypassing the
lantern.  For the 61 core lantern two different comparison fibres were
used, both with \SI{105}{\micro\metre} core diameter, nominal NA of
0.22, length of \SI{1}{\metre} and SMA connectors at both ends.  
The first data set used an Ocean Optics fibre patch cable while the
second used a Thorlabs M15L01 fibre patch cable.

The rotary stage was used to change the angle of incidence of the beam
in intervals of \SI{0.5}{\degree} for the lanterns
and \SI{1}{\degree} for the comparison fibres.  At each angle of
incidence the near infrared camera was used to take
images of the near field output of the photonic lantern/fibre.  The range of angles was
chosen so that the output flux fell to undetectable levels on both
sides of centre.  For the photonic lanterns the rotation of the stage
was reversed after reaching the end of the range of angles and a
second set of images were obtained for all angles in the range.  This
was done to enable a check of the repeatability of the measurements.
Sequences of dark frames were also obtained before and after each
sequence of near field images.

\subsection{Results}

\begin{figure}
\center\includegraphics[height=0.90\textheight]{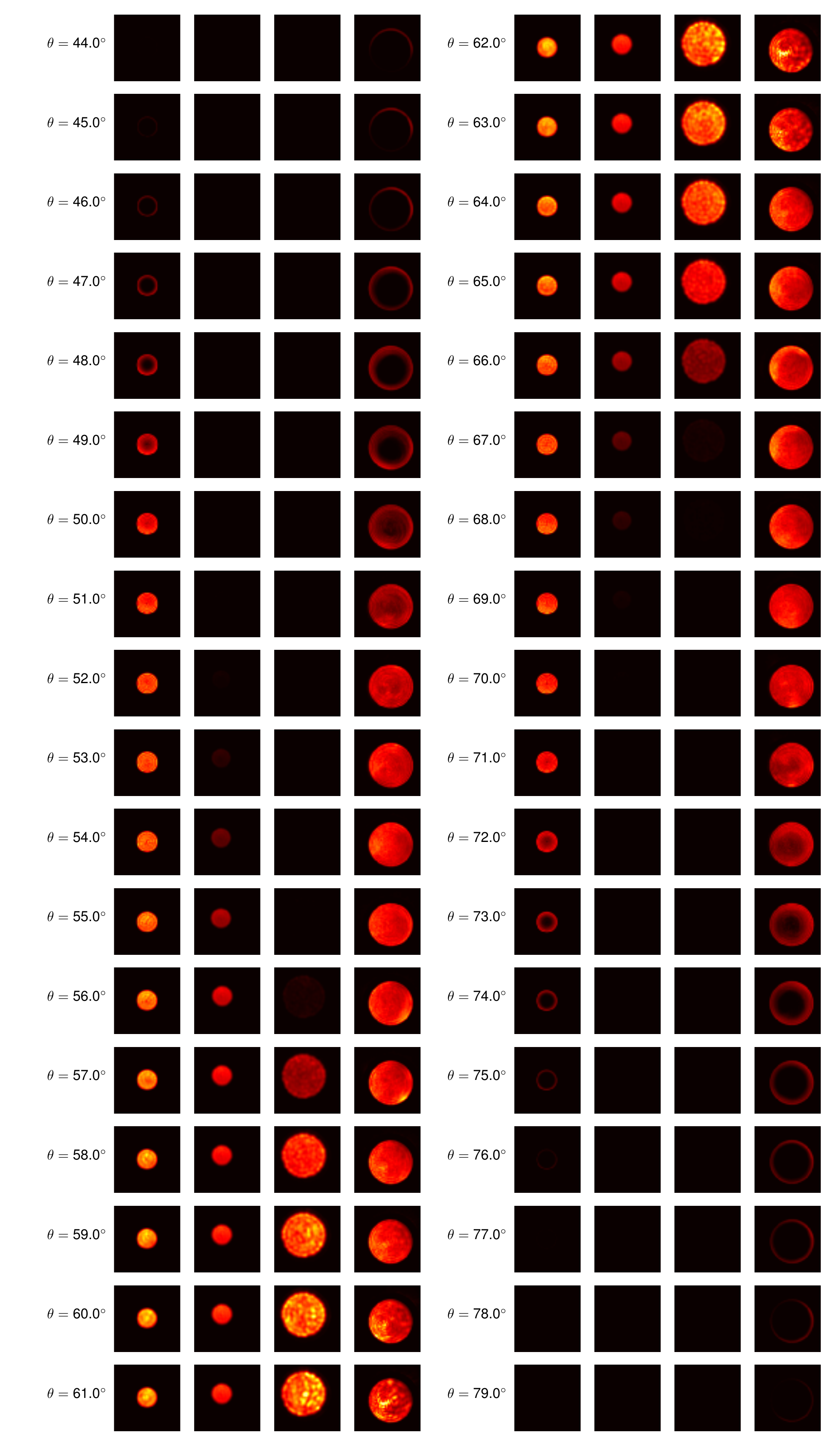}
\caption{\label{fig:nearfields}
Near field images of the output of the photonic lanterns and
comparison fibres.  From left to right the columns are from
the \SI{50}{\micro\metre} core comparison fibre, the GNOSIS 19 core
photonic lantern, the PL-13-12 61 core photonic lantern and the
\SI{105}{\micro\metre} core comparison fibre.  The angles refer to the
position of the rotary stage which has an arbitrary offset from the
angle of incidence.  Each image is a $150\times150$ pixel region from
a single exposure. The images have been dark subtracted but not flat
fielded.}
\end{figure}

Example near field output images of both photonic lanterns and two of
the comparison fibres are shown in figure \ref{fig:nearfields}.  These
images qualitatively illustrate some of the differences in behaviour between
a photonic lantern and an ordinary fibre.  Both of the comparison
fibres show large variations of the near field illumination patterns
with angle of incidence, at low angles of incidence variable modal
speckle patterns can be seen (the origin of fibre modal noise) while at
larger angles the output light becomes confined to increasingly narrow
rings of skew rays spiralling around the core-clad interface.  The 61
core PL-13-12 photonic lantern exhibits less change to the overall
illumination distribution as the angle of incidence changes, we see variable
modal speckle patterns but no change to the overall radial
distribution (i.e.\ no skew rays).  The 19 core GNOSIS lantern has a
highly consistent and uniform near field distribution, the combination
of a photonic lantern with sections of conventional multimode fibre on
input and output appears to effectively homogenise the output
regardless of the angle of incidence of the input.  These observations
hint at the potential utility of photonic lanterns for `fibre
scrambling', an application mentioned in section \ref{sec:intro}.

For the purposes of modelling the performance of an instrument using
photonic lanterns we are primarily interested in the throughput of the
lantern as a function of input angle.  We estimate the throughput by
dividing the total flux in the near field output images from the
photonic lanterns by the total flux at the same angle of incidence from the
comparison fibres.  This approach was adopted due to difficulties in
directly measuring the input flux because of interference and scattering effects.

\subsubsection*{GNOSIS lantern}


\begin{figure}
\center\includegraphics[width=0.6\textwidth]{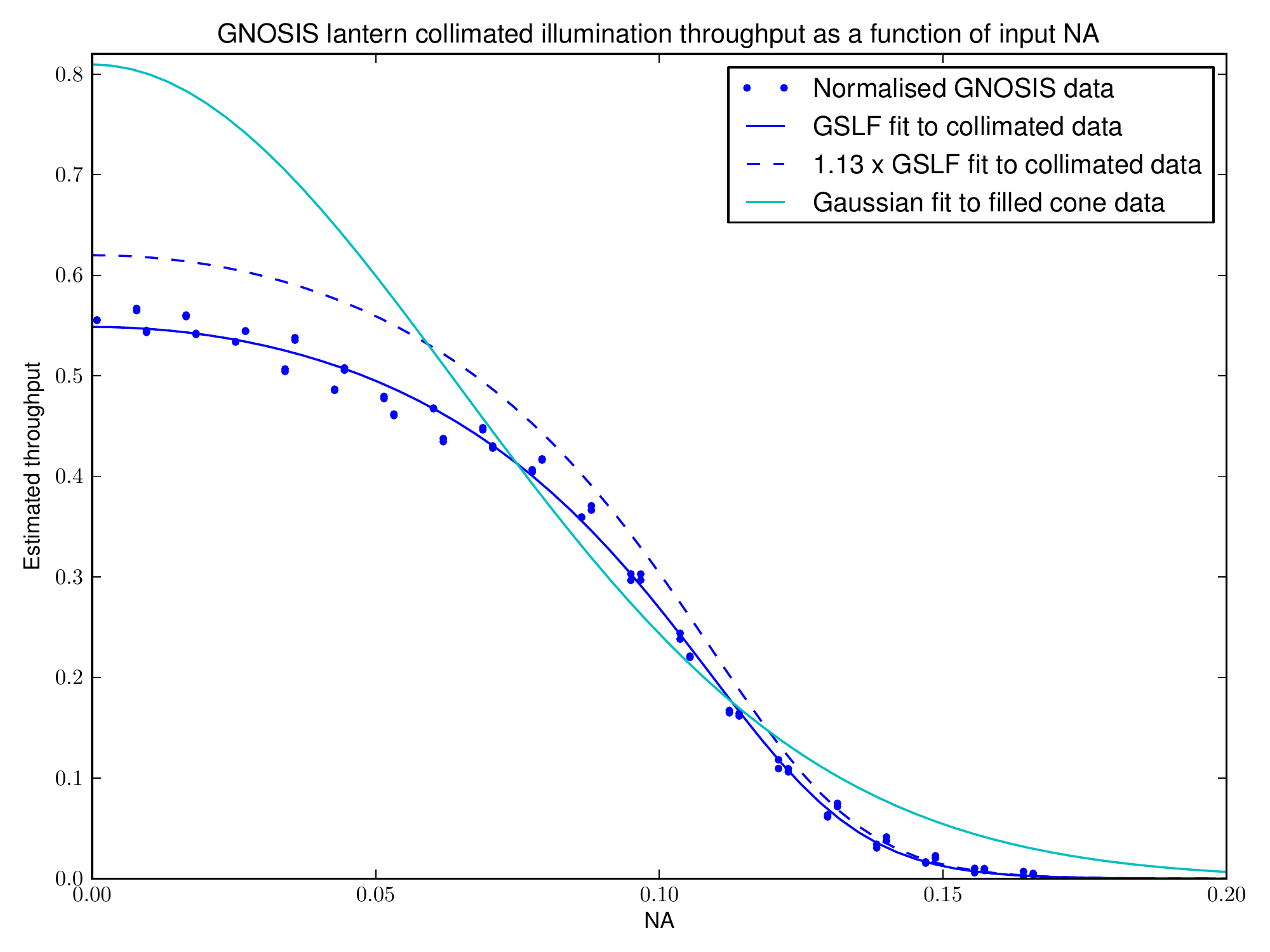}
\caption{\label{fig:g_fit}
Throughput of the GNOSIS photonic lantern with Thorlabs M16L01 SMA/FC
fibres relative to the M16L01 fibres alone. The throughput is for collimated
illumination and is plotted as a function of the numerical aperture corresponding to
the angle of incidence of the input illumination.
}
\end{figure}

\begin{figure}
\center\includegraphics[width=0.6\textwidth]{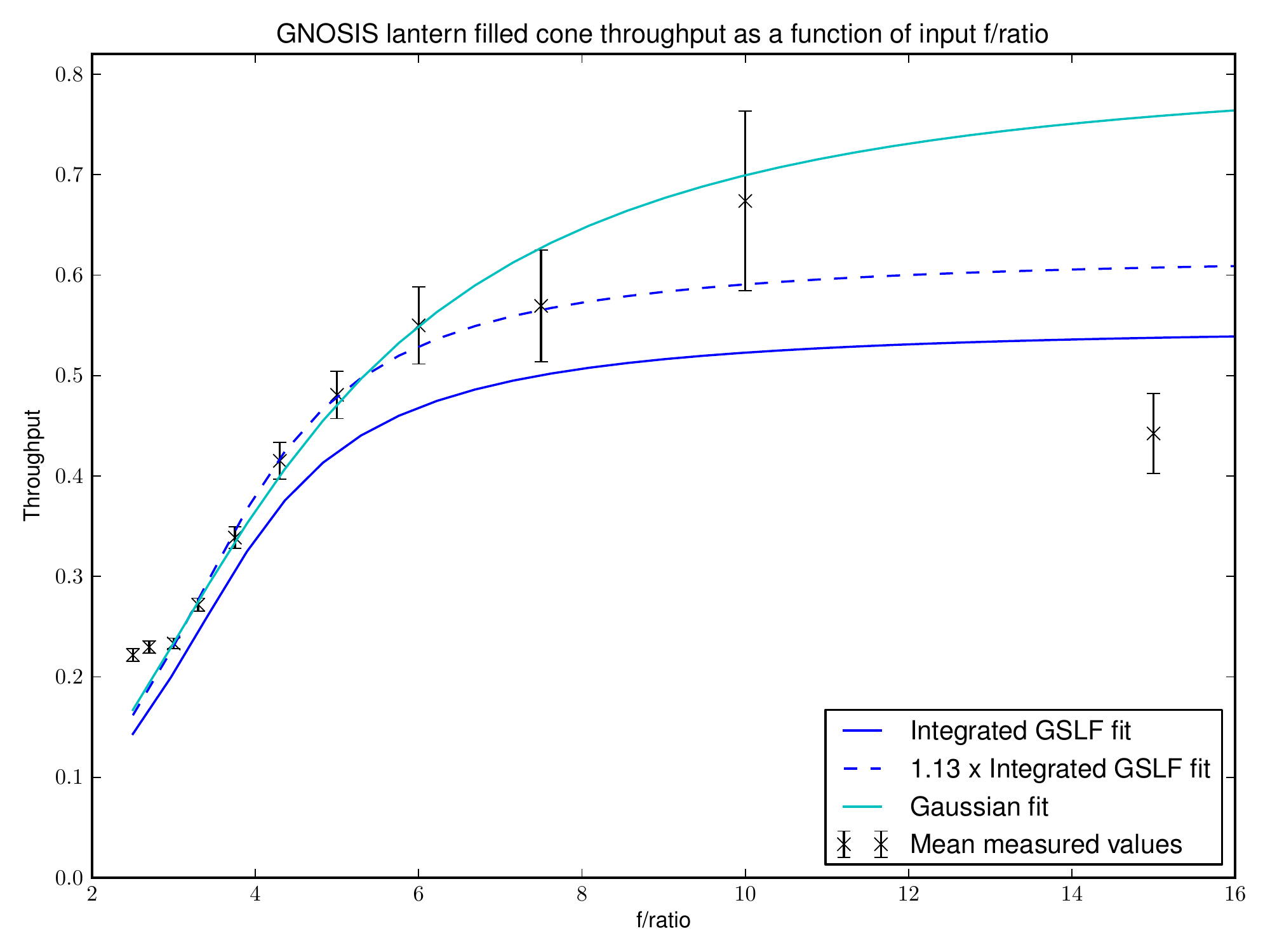}
\caption{\label{fig:g_int}
Measured throughput of the GNOSIS photonic lanterns when illuminated
by converging beams of various focal ratios.  The data shown are the
mean values at each focal ratio for the 7 channels of the GNOSIS
grating unit and are plotted together with the standard error on those
means.  Also shown are the throughput as a function of focal ratio
derived from integration of the GLSF fit to the collimated
illumination data as shown in figure \ref{fig:g_fit} and the best fit
to the data using a Gaussian double integral fitting function.
}
\end{figure}

To obtain throughput estimates the total output flux data
for the GNOSIS photonic lantern with the Thorlabs M16L01 fibre
patch cables were divided by a spline
fit to the corresponding data for the M16L01 fibre patch
cables alone.  The throughputs were then folded about \SI{0}{\degree} angle
of incidence and the angles converted to equivalent numerical apertures.  The
results are shown in figure \ref{fig:g_fit}. 

It was found that the throughput curve could
be well fit by a function of the form

\begin{equation}
\label{eqn:glsf}
\tau(\textrm{NA}) = \frac{A}{A_0}\left(1 - \left(1 + ae^{-(\textrm{NA}^2 -
      \textrm{NA}_0^2)/w^2}\right)^{-1/b}\right),
\end{equation}

where $\tau$ is the throughput, NA is the numerical aperture
($n\sin\theta$), $A$ is the peak throughput, NA$_0$, $w$, $a$ and $b$
are the other fitting parameters and

\begin{equation}
\label{eqn:a0}
A_0 = \left(1 - \left(1 + ae^{-\textrm{NA}_0^2/w^2}\right)^{-1/b}\right).
\end{equation}

We refer to this function as the Generalised Squared Logistic Function
(GSLF) by analogy with the similar Generalised Logistic Function.  The
GSLF fit to the GNOSIS photonic lantern throughput data is plotted in
figure \ref{fig:g_fit} and can be seen to be a good fit.  Also shown
in figure \ref{fig:g_fit} is the best fit Gaussian from the filled
cone measurements of section \ref{sec:filledcone} which clearly
deviates significantly from these data at both low and high NAs.  The
throughput curve as inferred from the collimated illumination data is
less strongly peaked and has narrower wings than the Gaussian but is not
entirely flat topped or steep sided, i.e.\ as expected the behaviour
of the photonic lantern is intermediate between that of a single mode
fibre and a highly multimoded fibre.

The discrepancy between the Gaussian fit to the filled cone data and
the collimated illumination data seen in figure \ref{fig:g_fit} does
not imply that the two sets of data are contradictory, rather it is a
consequence of the poor constraints on the high and low NA throughput
provided by the filled cone data.  In order to test the consistency of
the two data sets we performed numerical integrations of the GLSF fit
to convert the results from collimated illumination throughput as a
function of NA to filled cone illumination throughout as a function of
focal ratio.  We plot the resulting throughput curve together with the
experimental filled cone data and corresponding Gaussian fit in figure
\ref{fig:g_int}.  It can been seen that the unadjusted integrated GLSF fit is
inconsistent with the filled cone data but if the throughput is scaled
up by a constant factor of 1.13 the integrated GLSF fits the data
at least as well as the Gaussian fit, when the error bars and suspected
systematic errors are taken into account.  The need to scale up the
GLSF fit is unsurprising, as noted previously the GNOSIS channel used
to obtain the collimated illumination data is believed to have the
lowest overall throughput of the 7 in the grating unit while the
filled cone data is based on mean values from measurements of all
7 channels.  We conclude that both sets of data are consistent with
each other and with the GLSF model and that while the Gaussian model is
consistent with the filled cone data it is ruled out by the greater
discriminating power of the collimated illumination data.

\subsubsection*{61 core lantern}


\begin{figure}
\center\includegraphics[width=0.6\textwidth]{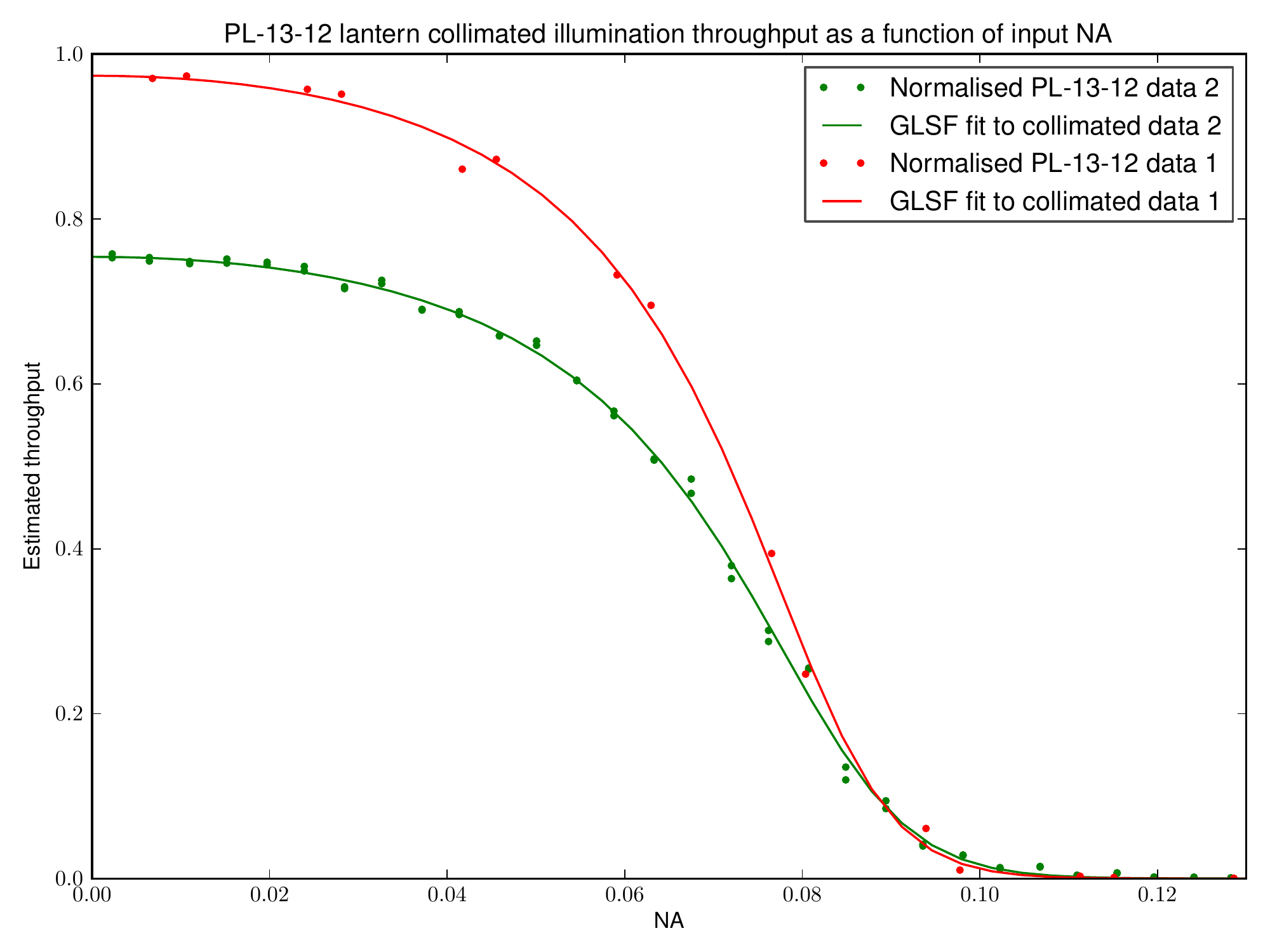}
\caption{\label{fig:p_fit}
Throughput of the bare PL-13-12 61 core photonic lantern relative to
the Thorlabs M15L01 (data 2) and Ocean Optics (data 1) comparison
fibres. The throughput is for collimated illumination and is plotted as
a function of the numerical aperture corresponding to the angle of
incidence of the input illumination. 
}
\end{figure}

The total flux value data for the PL-13-12 61 core photonic lantern
were processed in the same way as for the GNOSIS lantern.  The
resulting throughput estimates are shown in figure \ref{fig:p_fit}.  Two sets of data from the
are shown, an initial set with measurements taken at
\SI{1}{\degree} intervals in a single pass and using the Ocean Optics
comparison fibre, and a second set with measurements taken at \SI{0.5}{\degree}
intervals with two passes and using the Thorlabs M15L01 comparison
fibre. Also shown in figure \ref{fig:p_fit} are GLSF fits
to the two data sets, both are good fits.  The throughput curves from
the two data sets have very similar shapes but different normalisation
which suggests a problem with the use of the comparison fibres for
calibration.  The approach used for the PL-13-12 lantern where bare
lantern measurements are compared with a fibre patch cable is clearly
more prone to error than that used for the GNOSIS lantern where the lantern was
measured via fibre patch cables on input and output and the comparison
was with the same two fibre patch cables directly connected to one another.

\subsection{Analysis}

\begin{figure}
\center\includegraphics[width=0.6\textwidth]{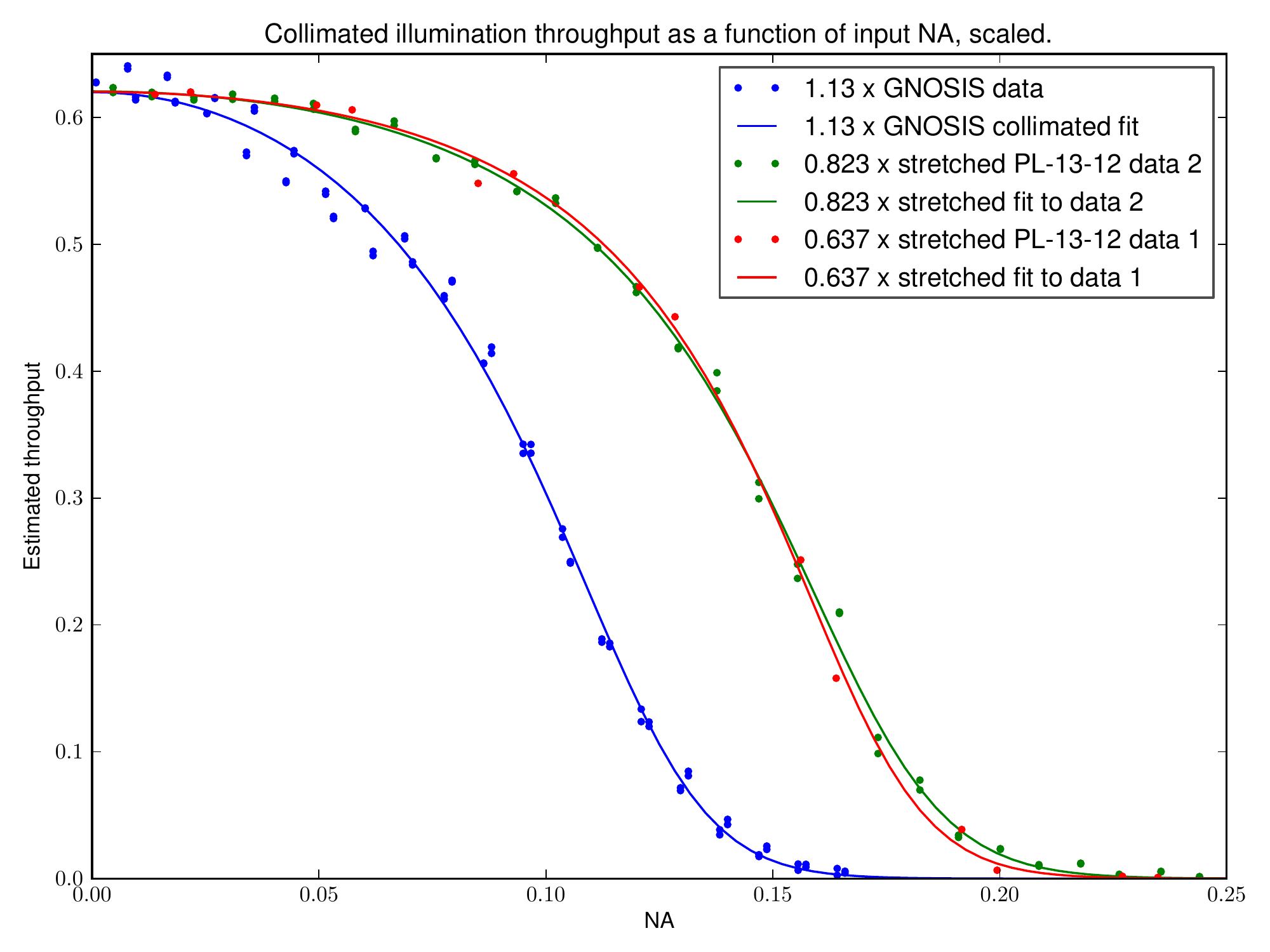}
\caption{\label{fig:all_scaled}
Comparison plot of all collimated illumination throughput against
input numerical aperture data together with GLSF fits, after rescaling to the same MMF output
core diameter and peak throughput.  The overall normalisation is
chosen to match the integrated throughput of the GNOSIS collimated
illumination data to the GNOSIS filled cone illumination throughput.}
\end{figure}

Figure \ref{fig:all_scaled} shows a comparison of all of the
collimated illumination data.  In this plot the numerical aperture
scale of the PL-13-12 data has been stretched by a factor of $105/50$
to compensate for the different fibre core size of that lantern's
multimode output (see equation \ref{eqn:napl}).  In normalising the
plots we have, in the absence of better data, considered the GNOSIS filled cone results to be
definitive, i.e.\ we have scaled the GNOSIS collimated illumination
data up by a factor of 1.13 and scaled both of the PL-13-12 data sets
so that all three sets of data have the same peak value.  Note that
the result is a conservative estimate of the throughput of a
photonic lantern, the GNOSIS measurements that have been used for the
normalisation are of an assembled FBG OH suppression unit and so
incorporate losses associated with the FBGs, MMF pigtails and multiple
fibre splices as well as the lantern itself.  Once normalised it is
apparent that the first PL-13-12 data set follows an essentially
identical curve to the second PL-31-12 data set so in the subsequent
analysis we only use the fit to the second, better sampled data set.


\begin{figure}
\center\includegraphics[width=0.6\textwidth]{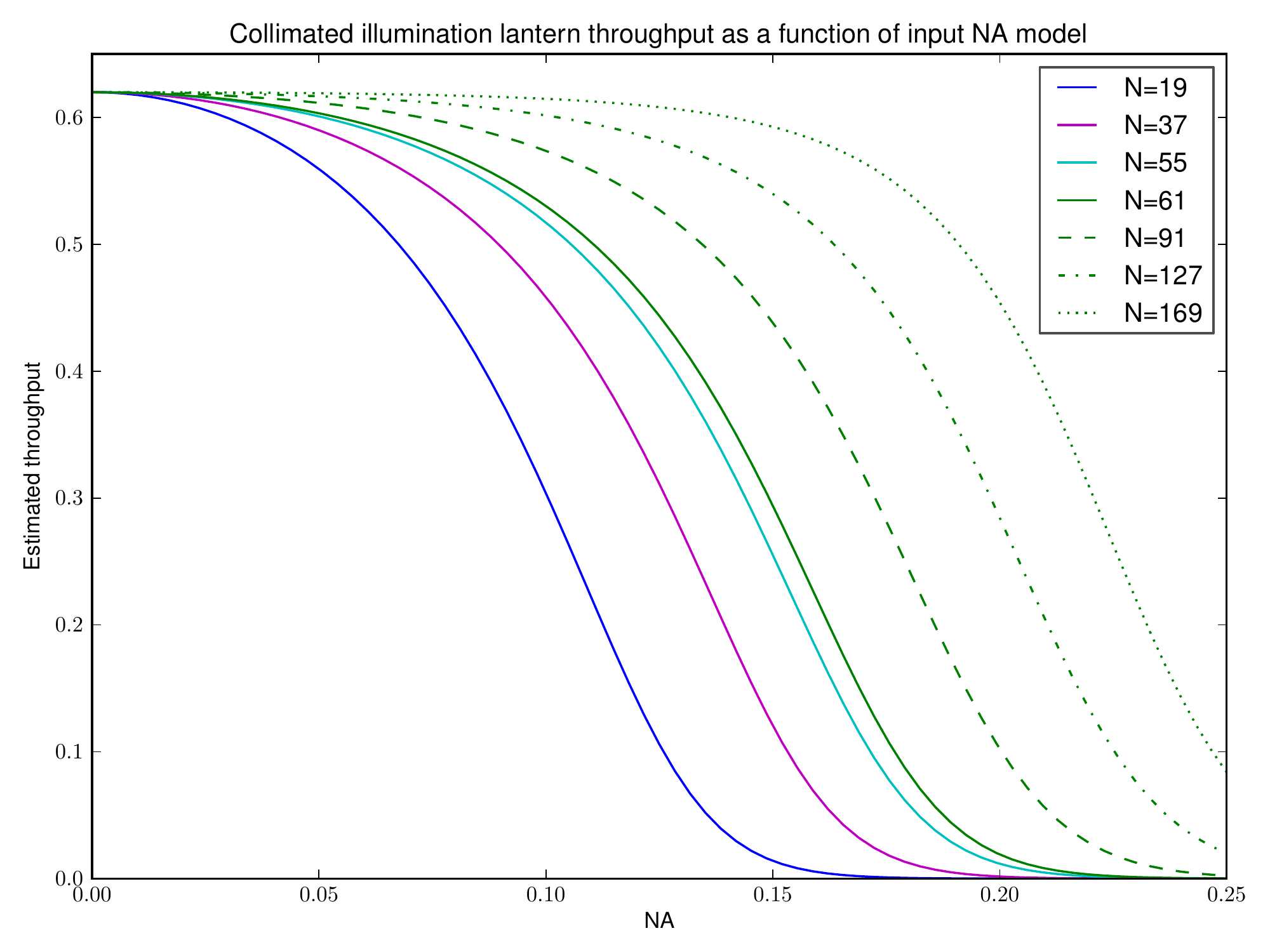}
\caption{\label{fig:model1}
Collimated illumination throughput against input numerical apeture for
the empirical photonic lantern model with
$\lambda=\SI{1.532}{\micron}$, $d=\SI{50}{\micron}$ and a range of
values of $N_\textrm{core}$.}
\end{figure}

Using the rescaled GNOSIS and PL-13-12 GLSF fits shown in figure
\ref{fig:all_scaled} we then constructed an empirical model for a
general photonic lantern.  The empirical model uses the GLSF of
equations \ref{eqn:glsf} and \ref{eqn:a0} with the peak throughput
parameter $A$ equal to that in figure \ref{fig:all_scaled} (0.62), 
the shape parameters $a$ and $b$ interpolated/extrapolated from the
rescaled GNOSIS and PL-13-12 values in proportion to
$\sqrt{N_\textrm{core}}$, and the NA scaling parameter 
$w$ and NA width parameter $\textrm{NA}_0$ interpolated/extrapolated
in the same way and then multiplied by
$\left(\lambda/\SI{1.532}{\micron}\right)\left(\SI{50}{\micron}/d\right)$
to rescale the NA axis in accordance with equation \ref{eqn:napl}.
Figure \ref{fig:model1} shows the resulting model for
$\lambda=\SI{1.532}{\micron}$, $d=\SI{50}{\micron}$ and
$N_\textrm{core}$ equal to each of the centred hexagonal numbers from
19 to 169 and 55.  We note that the shape of the throughput function
evolves only gradually with $N_\textrm{core}$, the primary effect of
increasing $N_\textrm{core}$ is an overall stretch in the NA axis. Even at
$N_\textrm{core}=169$ there is no sharp cut off, the behaviour of the
model photonic lantern still differs from that of a highly multimoded MMF.


\section{PRAXIS performance model}

As an illustration of the application of the empirical photonic
lantern model to astrophotonic instrument design we will use the PRAXIS
instrument as a case study.  PRAXIS is a near infrared (NIR)
spectrograph intended to serve as a testbed for FBG OH
suppression\cite{Horton2012,Content2014}.  It is a successor to the
GNOSIS instrument which successfully demonstrated the suppression of
OH lines by FBGs but was unable to produce a robust measurement of the
resulting interline sky background due to low throughput and
relatively high instrumental noise levels\cite{Ellis2012,Trinh2013a}.
PRAXIS will improve upon the sensitivity of GNOSIS and confirm whether
the reduction in interline background expected from FBG OH
suppression\cite{Ellis2008} occurs, and to what extent.  In addition
to accurate measurements of the sky background PRAXIS is intended to quantify the
practical benefits of FBG OH suppression by undertaking observations of
faint science targets.  Consequently the design of PRAXIS should be
optimised to produce the highest possible sensitivity to extended
sources (such as the sky) without significantly compromising
sensitivity to compact sources (e.g.\ low mass stars, high redshift
galaxies).  As the trade off between extended and compact source
sensitivity is primarily determined by an instrument's on sky
sampling/field of view it is clear that any attempt to determine the
optimum design parameters for PRAXIS must take into account the
angular dependence of photonic lantern transmission.  We
have developed a instrument performance model of PRAXIS to enable us
to predict its sensitivity and determine the optimum values of design
parameters.

The PRAXIS spectrograph is fed by 19 optical fibres, 7 of
which have photonic lanterns containing OH suppression FBGs.  The fibres are
illuminated by an integral field unit (IFU) which consists of a close
packed hexagonal microlens array (MLA).  The central 7 microlenses
project telescope pupil images onto the 7 OH suppressed fibres, together
these 7 microlens form the main entrance aperture of the
instrument.  The remaining 12 fibres are fed by the ring of microlens
around the central 7, these fibres will be used primarily for comparison
with the OH suppressed fibres.  The IFU is itself illuminated with a
magnified image of the sky by a fore optics unit which also includes a
cold stop.   The PRAXIS performance model is not a full end to end systems
engineering model based on detailed optical simulation, instead it is a
simplified model intended to enable efficient sampling of the design
parameter space without requiring full optical design details.  The model takes
into account:
\begin{itemize}
  \item{Atmospheric seeing}
  \item{Telescope throughput}
  \item{Telescope entrance pupil}
  \item{Telescope point spread function (PSF)}
  \item{Fore optics throughput}
  \item{Microlens apertures}
  \item{Diffraction from microlens apertures}
  \item{Geometric FRD from microlens non-telecentricity}
  \item{Photonic lantern throughput as a function of angle of
      incidence}
  \item{Spectrograph throughput}
  \item{Spectral resolution and spectrograph PSF size}
  \item{Instrument thermal background}
  \item{Detector quantum efficiency (QE), dark current and read out
      noise}
\end{itemize}
The model does not take into account optical aberrations, optical
misalignments, telescope pointing errors, cosmic rays or the effects of optimal
spectral extraction and weighted combination of spectra, however all
of these factors are expected to have small or minimal effects on the
performance of PRAXIS.  The PRAXIS performance model has been used to
investigate and derive specifications for numerous design and performance parameters
including $N_\textrm{core}$, pupil image magnification at the fibre
inputs, microlens non-telecentricity (insignificant), detector dark
current, detector read noise and overall instrument thermal
background.  The photonic lantern empirical model was also used to place an
upper limit on the thermal emissivity of the lanterns which in turn
lead to a decision to cool the OH suppression unit.  For the purposes
of this case study we will concentrate on the results for varying
on-sky sampling/field of view as these are the most directly
influenced by the photonic lantern model.


\begin{figure}
\center\includegraphics[width=0.9\textwidth]{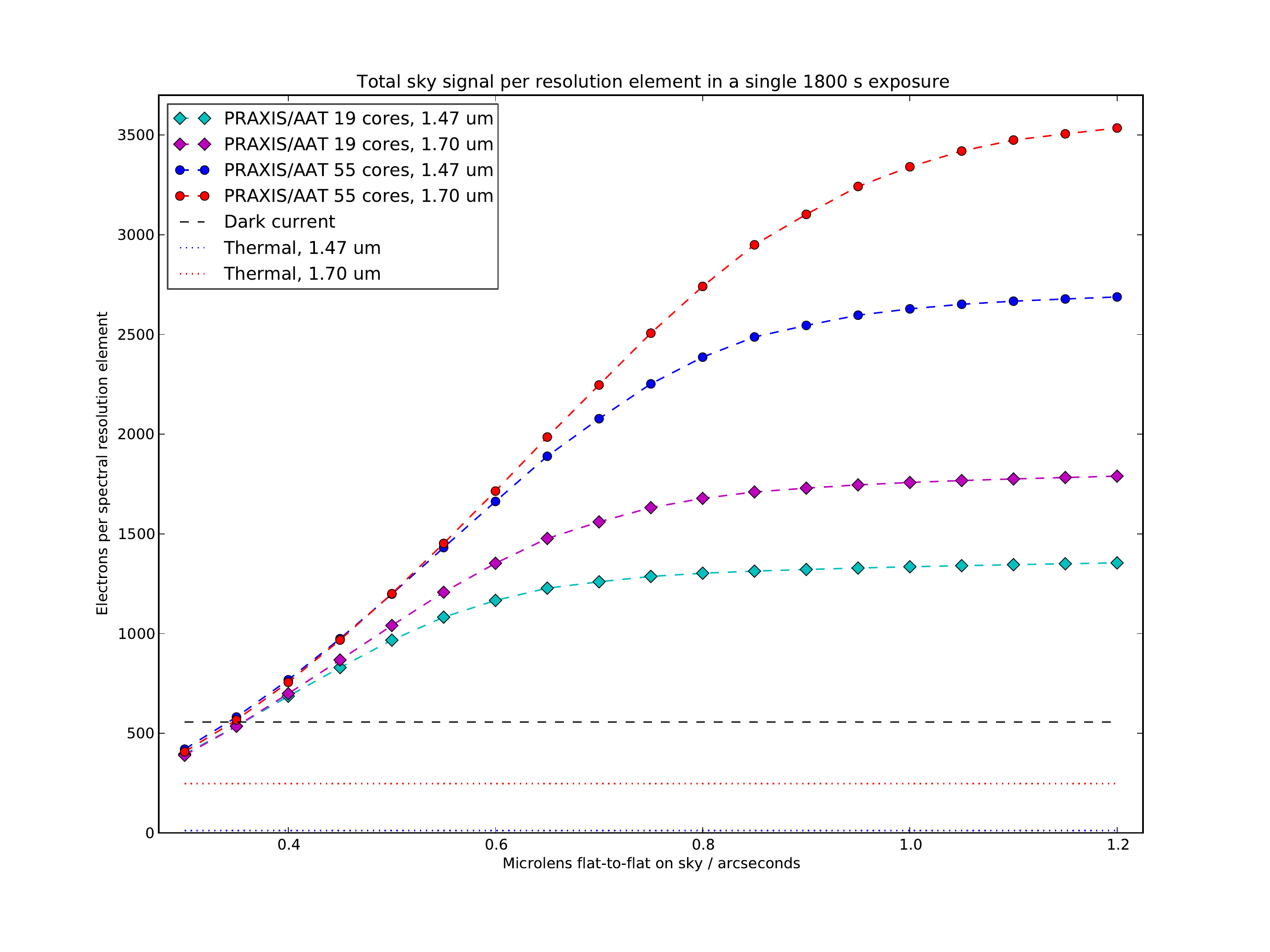}
\caption{\label{fig:sky_counts}
Predicted detected electrons per spectral resolution
element for the summed sky spectra of an array of 7 photonic lanterns fed
by hexagonal microlenses versus the on sky flat-to-flat field of view
per microlens.  The spectral resolution is $R=2500$, the exposure time
is \SI{1800}{\second} and the sky surface brightness is assumed to be \SI{500}{\psb}}
\end{figure}


\begin{figure}
\center\includegraphics[width=0.9\textwidth]{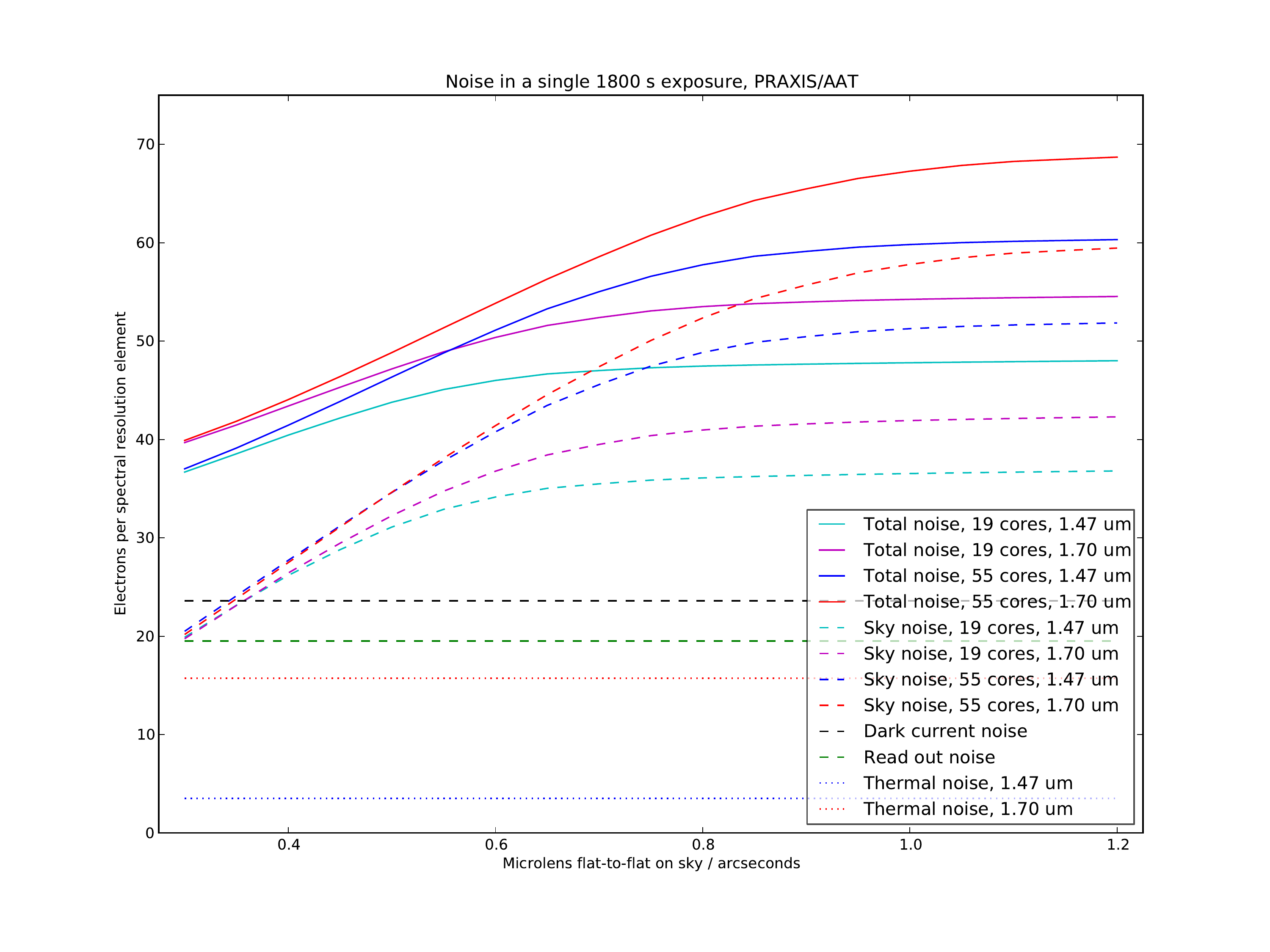}
\caption{\label{fig:sky_noise}
Predicted noise electrons per spectral resolution
element for the summed sky spectra of an array of 7 photonic lanterns fed
by hexagonal microlenses versus the on sky flat-to-flat field of view
per microlens.  The spectral resolution is $R=2500$, the exposure time
is \SI{1800}{\second} and the sky surface brightness is assumed to be \SI{500}{\psb}}
\end{figure}

\begin{figure}
\center\includegraphics[width=0.9\textwidth]{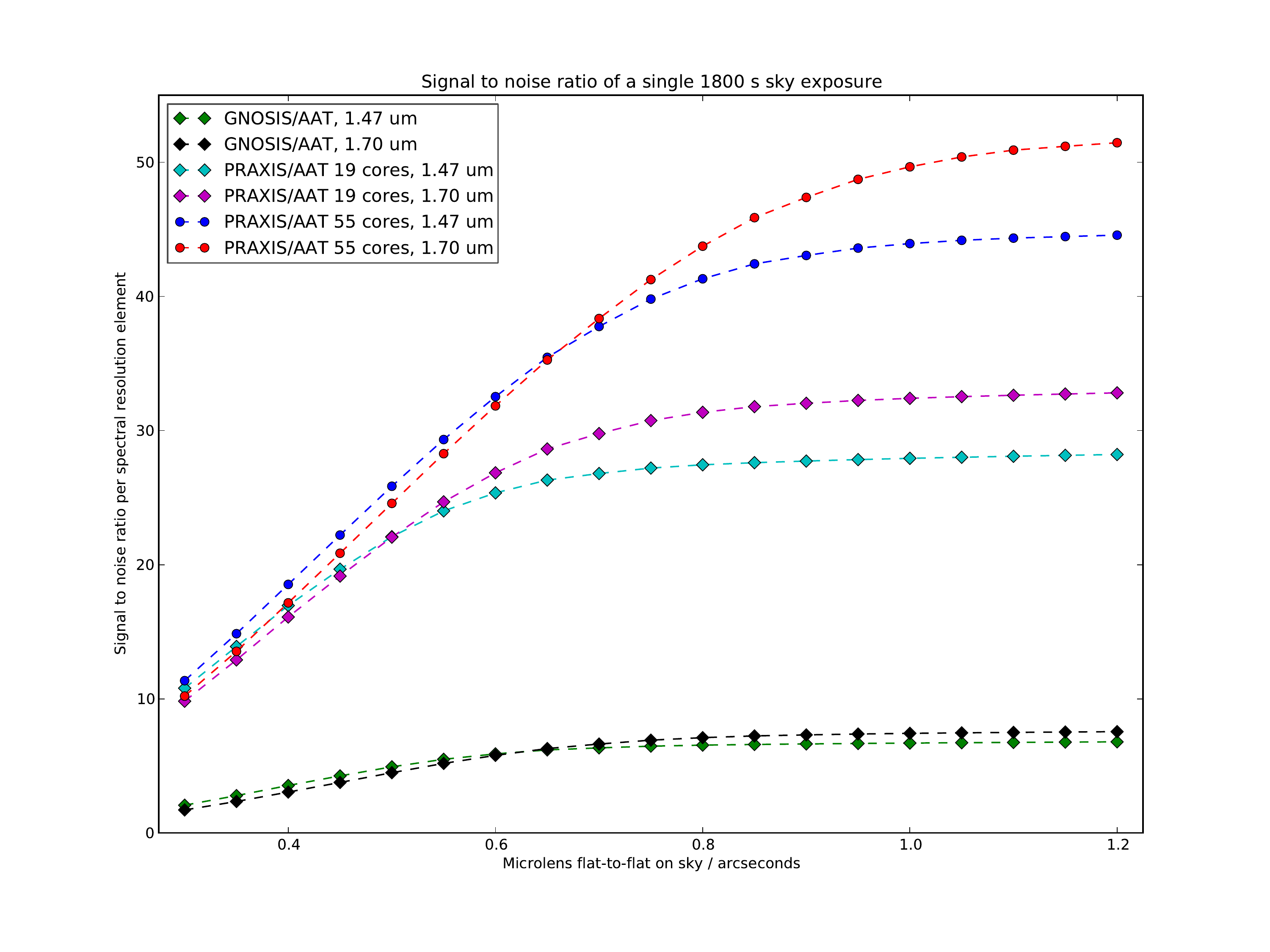}
\caption{\label{fig:sky_snr}
Predicted signal to noise ratio per spectral resolution
element for the summed sky spectra of an array of 7 photonic lanterns fed
by hexagonal microlenses versus the on sky flat-to-flat field of view
per microlens.  The spectral resolution is $R=2500$, the exposure time
is \SI{1800}{\second} and the sky surface brightness is assumed to be
\SI{500}{\psb}}
\end{figure}



\begin{figure}
\center\includegraphics[width=0.9\textwidth]{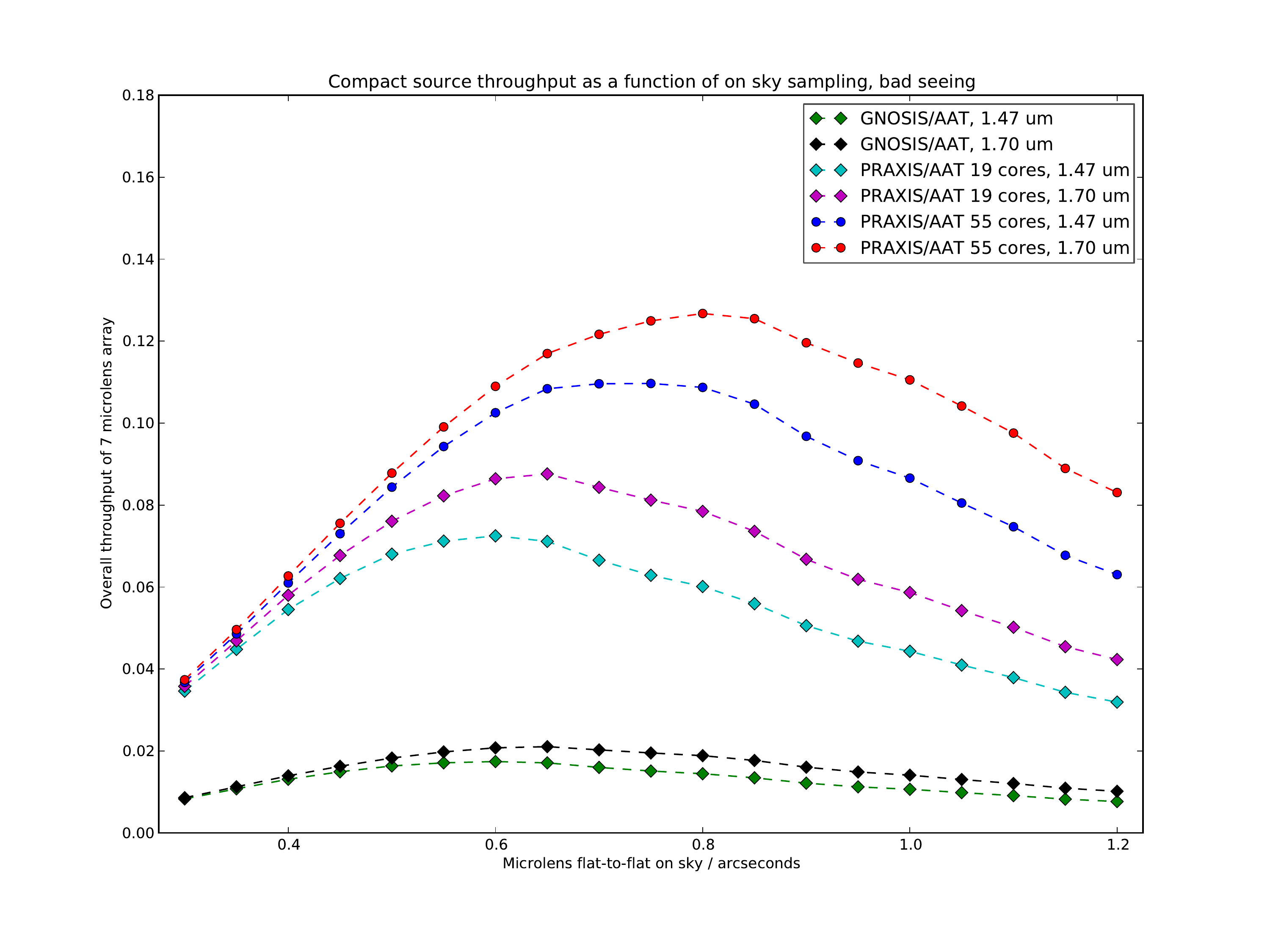}
\caption{\label{fig:cs_throughput}
Overall system throughput for compact source centred on an
array of 7 photonic lanterns fed by hexagonal microlenses versus the
on sky flat-to-flat field of view per microlens.  The atmospheric
seeing is assumed to be \ang{;;1.67} at $\lambda=\SI{500}{\nano\metre}$ which equates
to \ang{;;1.33} at $\lambda=\SI{1.585}{\micron}$}
\end{figure}



\begin{figure}
\center\includegraphics[width=0.9\textwidth]{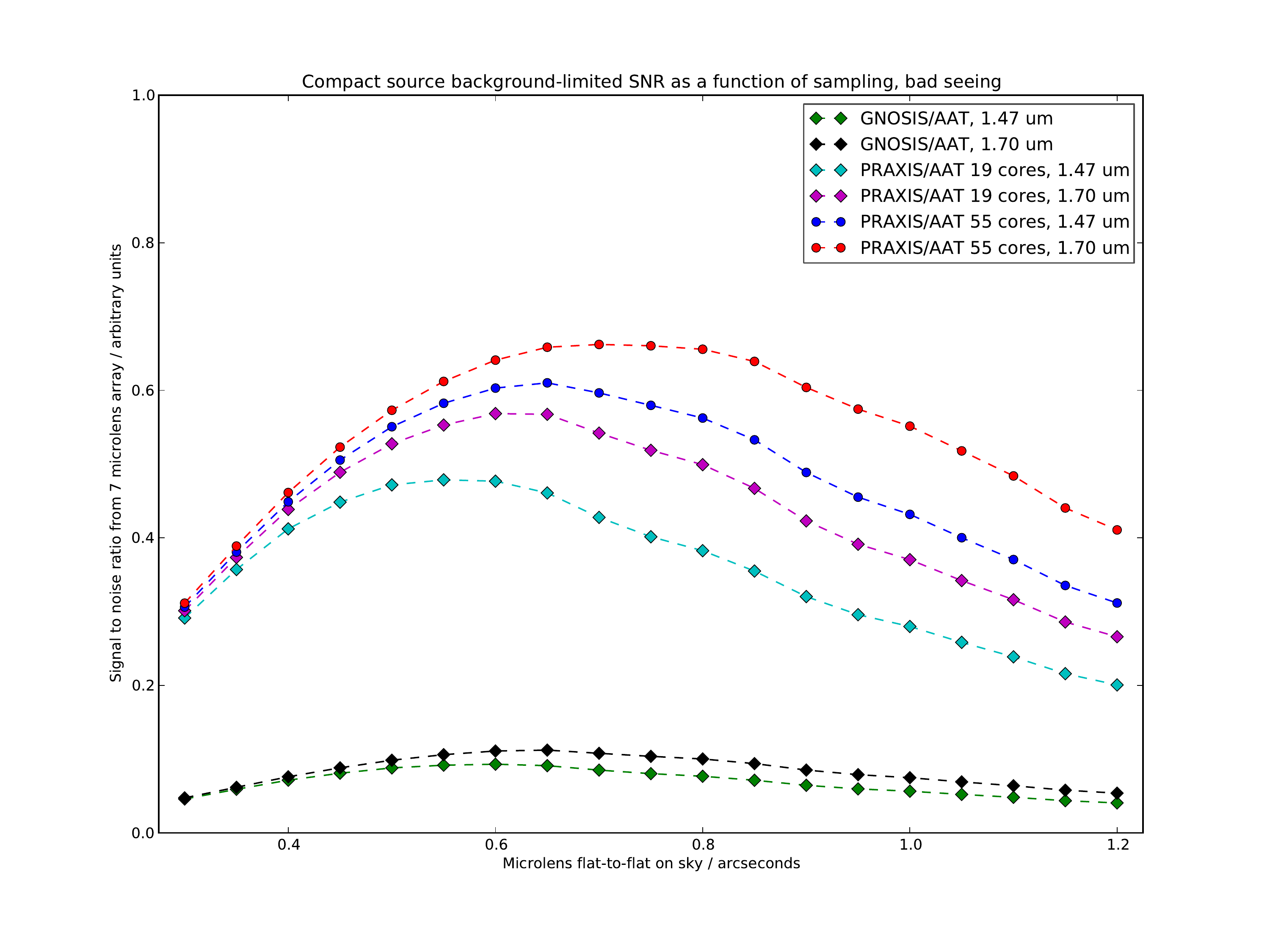}
\caption{\label{fig:cs_snr}
Predicted signal to noise ratio per spectral resolution element for
the summed spectra of a compact source centred on an
array of 7 photonic lanterns fed by hexagonal microlenses versus the
on sky flat-to-flat field of view per microlens.  The source is
assumed to be fainter than the background, the spectral resolution is $R=2500$, the exposure time
is \SI{1800}{\second}, the sky surface brightness is assumed to be
\SI{500}{\psb} and the atmospheric
seeing is assumed to be \ang{;;1.67} at $\lambda=\SI{500}{\nano\metre}$ which equates
to \ang{;;1.33} at $\lambda=\SI{1.585}{\micron}$}
\end{figure}

As the first purpose of PRAXIS is an accurate measurement of the sky
background we will look first at the predicted sky counts for a
typical single exposure of \SI{1800}{\second}.  Figure \ref{fig:sky_counts} shows the
total detected electrons per spectral resolution element in the sky
spectrum obtained by summing the signal from the 7 central fibres as a
function of the on sky sampling/field of view per microlens.  The
calculations have been made for the \SI{3.9}{\metre} Anglo-Australian Telescope (AAT)
which it where the commissioning and initial use of PRAXIS will take
place.  Points
have been calculated for two different values of $N_\textrm{core}$,
$N_\textrm{core}=19$ corresponds to the initial implementation of
PRAXIS which will use the existing photonic lanterns that were built
for GNOSIS while $N_\textrm{core}=55$ represents a planned upgrade of
PRAXIS to use new photonic lanterns manufactured from multicore
fibre\cite{Min2012,Haynes2012}.  The calculations have also been
performed for both ends of the PRAXIS wavelength range,
$\lambda=\SI{1.47}{\micron}$ and $\lambda=\SI{1.70}{\micron}$.  The
model assumes a sky surface brightness of \SI{500}{\psb} which is the
approximate interline sky brightness estimate obtained for Siding
Spring Observatory from GNOSIS data\cite{Trinh2013b}, and a spectral
resolution of $R=2500$.  The total sky counts increase as the field of
view increases before gradually levelling off.  The levelling off is
due to the drop off in transmission of the photonic lanterns for large
angles of incidence.  Figure \ref{fig:sky_counts} also
shows the current best estimates for the detector dark current
(\SI{9e-3}{e\tothe{-}\per\pix\per\second}) and instrument thermal background
levels (\SIrange[range-phrase=--,range-units=brackets]{0.2}{4.0e-3}{e\tothe{-}\per\pix\per\second}),
as desired these fall below the sky counts for all fields of view per
microlens $\gtrsim\ang{;;0.4}$. 

In figure \ref{fig:sky_noise} we look
at the contributions of the various noise sources included in the
model. The parameters and axes are the same as in figure
\ref{fig:sky_counts} and the plot shows the total noise per resolution
element along with the contributions from sky signal itself, the
detector dark current, detector read out noise
(\SI{3.33}{e\tothe{-}\per\pix\per\second} for multiple read mode) and
instrument thermal background.  We see that the Poisson noise from the
sky signal is the dominant source of noise for FoV per microlens
$\gtrsim\ang{;;0.4}$ but only by margins of $\sim$2--3 over the other
individual noise sources, it is clear that it is vitally important to
minimise all of the instrumental sources of noise in order to ensure
sky noise limited observations and so maximise sensitivity.  

By combining the results from figures \ref{fig:sky_counts} and
\ref{fig:sky_noise} we can calculate the predicted signal to noise
ratio (SNR) of the summed sky spectrum, this is plotted in figure
\ref{fig:sky_snr}. The SNR of PRAXIS with $N_\textrm{core}=19$ is
expected to reach up to $\sim30$ for a sky surface brightness of
\SI{500}{\psb}, for $N_\textrm{core}=55$ the SNR can reach
$\sim45$.  We have almost made the same calculations for an
$N_\textrm{core}=19$ instrument
with the lower throughput and higher instrumental noise of GNOSIS, for
this GNOSIS-like instrument the SNR does not exceed $\sim8$ and at the
FoV per microlens of \ang{;;0.4} as used by the actual GNOSIS
instrument the SNR in a single exposure is only $\sim2.5$.  These
results give us confidence that PRAXIS will be able to acheive its aim
of robust and accurate measurements of the interline sky background,
at least for sky surface brightness levels of $\sim$\SI{100}{\psb} or
more.  

In order to confirm the optimal Fov per
microlens we need to also consider observations of compact objects.
Figure~\ref{fig:cs_throughput} shows the calculated
overall throughput for observations of a compact source using the 7
central fibres.  The source is assumed to be point like such that the
observered size is determined purely by atmospheric seeing. We adopt a
Gaussian seeing disc with FWHM of \ang{;;1.67} at
$\lambda=\SI{500}{\nano\metre}$, which becomes \ang{;;1.33} at
$\lambda=\SI{1.585}{\micron}$.  This is considered
representative of seeing conditions at the AAT.  We note that contrary to
the extended source (sky) case there
is a value of the FoV per microlens which maximises the
received signal, and increasing the FoV per microlens further actually
reduces the overall throughput.  The reason for this is that the
effective entrance aperture of each microlens is limited by the
acceptance cone of the attached photonic lantern, consequently there
comes a point at which increasing the size of the microlenses further
simply opens up gaps in the effective entrance aperture of the
instrument as a whole thereby reducing the overall throughput. Peaks
in figure \ref{fig:cs_throughput} do not necessarily represent the
optimal values of FoV per microlens, though, because what we should be
optimising is SNR rather than throughput.  In principle the way SNR
varies depends on the ratio between the brightness of the source and
the background however we can make the reasonable assumption that the
majority of targets of scientific interest will be significantly
fainter than the background, i.e. the observations will be background
limited.  In this limit the SNR will be proportional to the ratio
between the compact source throughput and the background noise (i.e.\
the total noise from figure \ref{fig:sky_noise}).  We plot this SNR
parameter in figure \ref{fig:cs_snr}, note that the peaks have shifted
to slightly lower FoV per microlens because larger Fov results in
increased sky background noise.  Selecting the best FoV per microlens
is still not straightforward as there is a trade off between
sky/extended source sensitivity and compact source sensitivity but
figures \ref{fig:sky_snr} and \ref{fig:cs_snr} do at least provide
the information required for an informed decision.  The PRAXIS project
have selected \ang{;;0.55} and \ang{;;0.8} for $N_\textrm{core}=19$
and 55 respectively.

\section{Further work} 

The photonic lantern experiments undertaken so far have measured only
the total output flux (filled cone tests) or the near field output
light distribution (collimated illumation tests).  We intend to extend
our experimental study and empirical model to also include the far field output
light distribution as a function of input angle of incidence.  Being
able to reliably predict the angular distribution of light that will emerge
from a photonic lantern will be useful for the optimised design of
instrument components fed by photonic lanterns, e.g.\ multimode fibre
bundles and spectrographs.  By measuring both near and far field output
light distributions as a function of input illumination we will also
obtain valuable data for quantifying the `fibre scrambling'
effectiveness of the lanterns under test.

The measurements so far have also been restricted to only two values
of $N_\textrm{core}$, in fact only two individual lanterns for the
collimated illumination tests.  Both tested lanterns are of the
discrete single mode fibre type\cite{Noordegraaf2009} too, whereas photonic lanterns can also
be manufactured from multicore fibre\cite{Birks2012} or by the femtosecond
laser direct write process\cite{Spaleniak2013}.  We plan to increase
the parameter space covered by our results by testing 31 and 55 core multicore type photonic
lanterns in the near future.

\bibliography{/home/ajh/Documents/Papers/library}   
\bibliographystyle{spiebib}   

\end{document}